\documentclass{aa}
\usepackage{natbib,psfig,graphicx}
\begin{document}


\title{The build-up of the Coma cluster by infalling substructures}

\offprints{C. Adami: christophe.adami@oamp.fr}

\author{
C. Adami\inst{1}
\and A. Biviano\inst{2} 
\and F. Durret\inst{3,4} 
\and A. Mazure\inst{1}  
}
   
\institute{
Laboratoire d'Astropysique de Marseille, UMR 6110 CNRS-Universit\'e de
Provence, Traverse du Siphon, Les Trois Lucs, 13012 Marseille, France
\and
INAF / Osservatorio Astronomico di Trieste, via G.B. Tiepolo 11, 
34131, Trieste, Italy
\and
Institut d'Astrophysique de Paris, UMR 7095,Universite Pierre \& Marie Curie,
98 bis Bd Arago, 75014 Paris, France
\and
Observatoire de Paris, LERMA, 61 Av. de l'Observatoire, 75014 Paris, France
}

\date{Accepted . Received ; Draft printed: \today}

\authorrunning{Adami et al.}

\titlerunning{The build-up of the Coma cluster}

\abstract{We present a new multiwavelength analysis of the Coma
cluster subclustering based on recent X-ray data and on a compilation
of nearly 900 redshifts. We characterize subclustering using the Serna
\& Gerbal (1996) hierarchical method which makes use of galaxy
positions, redshifts, and magnitudes, and identify 17 groups. One of
these groups corresponds to the main cluster, one is the well known
group associated with the infalling galaxy NGC~4839 and one is
associated with NGC~4911/NGC~4926. About one third of the 17 groups
have velocity distributions centered on the velocities of the very
bright cluster galaxies they contain (magnitudes $R < 13$).

In order to search for additional substructures, we make use of the
isophotes of X-ray brightness residuals left after the subtraction of
the best-fit $\beta$-model from the overall X-ray gas distribution
(Neumann et al. 2003). We select galaxies within each of these
isophotes and compare their velocity distributions with that of the
whole cluster. We confirm in this way the two groups associated,
respectively, with NGC~4839, and with the southern part of the
extended western substructure visible in X-rays.

We discuss the group properties in the context of a scenario
in which Coma is built by the accretion of groups infalling from the
surrounding large scale structure. We estimate the recent mass
accretion rate of Coma and compare it with hierarchical models of
cluster evolution.


}

\maketitle

\section{Introduction}

Being located relatively nearby, Coma is one of the best
studied clusters of galaxies at almost all wavelengths.
For decades, it has been considered as the prototype of rich relaxed
clusters (e.g. Kent \& Gunn 1982).  However, following seminal
remarks or partial studies (e.g. Shane \& Wirtanen 1954, Quintana
1979, Valtonen \& Byrd 1979, Baier 1984, Perea et al. 1986 and
references in the review by Biviano 1998) the idea that Coma exhibits
a complex structure became commonly accepted with the results of
Fitchett \& Webster (1987) and Mellier et al. (1988, M88 hereafter)
in the late 80's.

Although the very first X-ray observations of the Coma Intra Cluster
Medium (ICM) already gave a hint of a possible complex central structure
(e.g. Tanaka et al. 1982),
the X-ray community recognized this fact only in 1993 (Davis \&
Mushostzky 1993).  Since then, new structures and properties are
regularly (re)discovered and better quantified.

Both optical and X-ray data lead to the present view where, in
particular, major substructures are present around NGC~4839, a group
which appears to be falling onto the Coma cluster core (e.g. Neumann
et al. 2001), and around the two main galaxies NGC~4874 and NGC~4889
(e.g. Gurzadyan \& Mazure 2001), the former possibly being the
detached core of a larger tidally stripped galaxy structure moving
northward (see e.g. Adami et al. 2005a).  Other substructures may be
present around NGC~4911 (see e.g. M88, Neumann et al. 2003, N03
hereafter). In the core, bright galaxies appear to be associated with
the two cluster dominant objects NGC~4874 and NGC~4889, while faint
galaxies have a smoother distribution (Biviano et al. 1996), but the
significance of this different distribution is not fully established
(Edwards et al. 2002). \L okas \& Mamon (2003) found that the 
early-type galaxy distribution in Coma is consistent with these
galaxies being on nearly isotropic orbits. Their result was consistent
with Adami et al.'s (1998) result, based on kinematical modelling,
that early-type Coma galaxies have moderately tangential
orbits. Additionally, Adami et al. also found moderately radial orbits
for the late-type galaxy population.

Due to the difficulty of collecting homogeneous velocity catalogues on
broad regions, structural and dynamical analyses of Coma on large
scales are generally based on photometric data (e.g. on the Godwin et
al. 1983 catalogue, hereafter GMP). Hence, up to now, the relatively
small number of redshifts available strongly penalized any kinematical
analysis.  For example, in order to locate and compute velocity
dispersions for typical group-size substructures ($\sim$100
h$^{-1}$kpc) in the dense cluster regions ($\sim$1h$^{-1}$Mpc), one
needs about 10 redshifts per 100$\times$100 h$^{-2}$kpc$^2$ ``pixel''
(e.g. Lax 1985), which therefore leads to a total number of redshifts of
about 1000. Such a number starts to be reachable by combining all
literature redshift catalogues for the Coma cluster. It is interesting
to note that, even in the era of modern multi-object spectrographs,
Coma is one of the few clusters (if not the only one) with such a
redshift sample available.

Several problems remain however. We must take into account the fact
that relaxation, survival and virialisation times depend on the galaxy
mass, morphological type, orbit, etc. In practice, we never have
access to all these parameters at the same time for all galaxies of a
given sample. It is therefore crucial to use another estimator of the
cluster dynamics: the hot gas X-ray emission. With the large
XMM-Newton field of view and collecting power, we now have access to
deep X-ray data over a large field for the Coma cluster (e.g. Neumann
et al. 2001).

The goal of this paper is to deepen our knowledge of the structure of
the Coma cluster and draw a scenario for its build-up through the
accretion of galaxy groups. For this we make use of statistical
descriptions of the galaxy velocity distributions and apply the
hierarchical method developed by Serna \& Gerbal (1996, hereafter SG)
combined with X-ray information, based on a large catalogue of
redshifts and recent XMM-Newton results (Neumann et al. 2001).  The
paper is organized as follows: our data-set is described in Section~2;
the methods and results of the substructure analysis are presented in
Section~3; our results are discussed in Section~4 and summarized in
Section~5. We assume a Hubble constant H$_0$=70 km~s$^{-1}$~Mpc$^{-1}$
throughout this paper. As for the center of the Coma cluster we take
$\alpha$=12h57.3mn, $\delta$=+28$^\circ$14.4' (equinox 1950.0; see
GMP).

\section{Samples}

\begin{figure}
\mbox{\psfig{figure=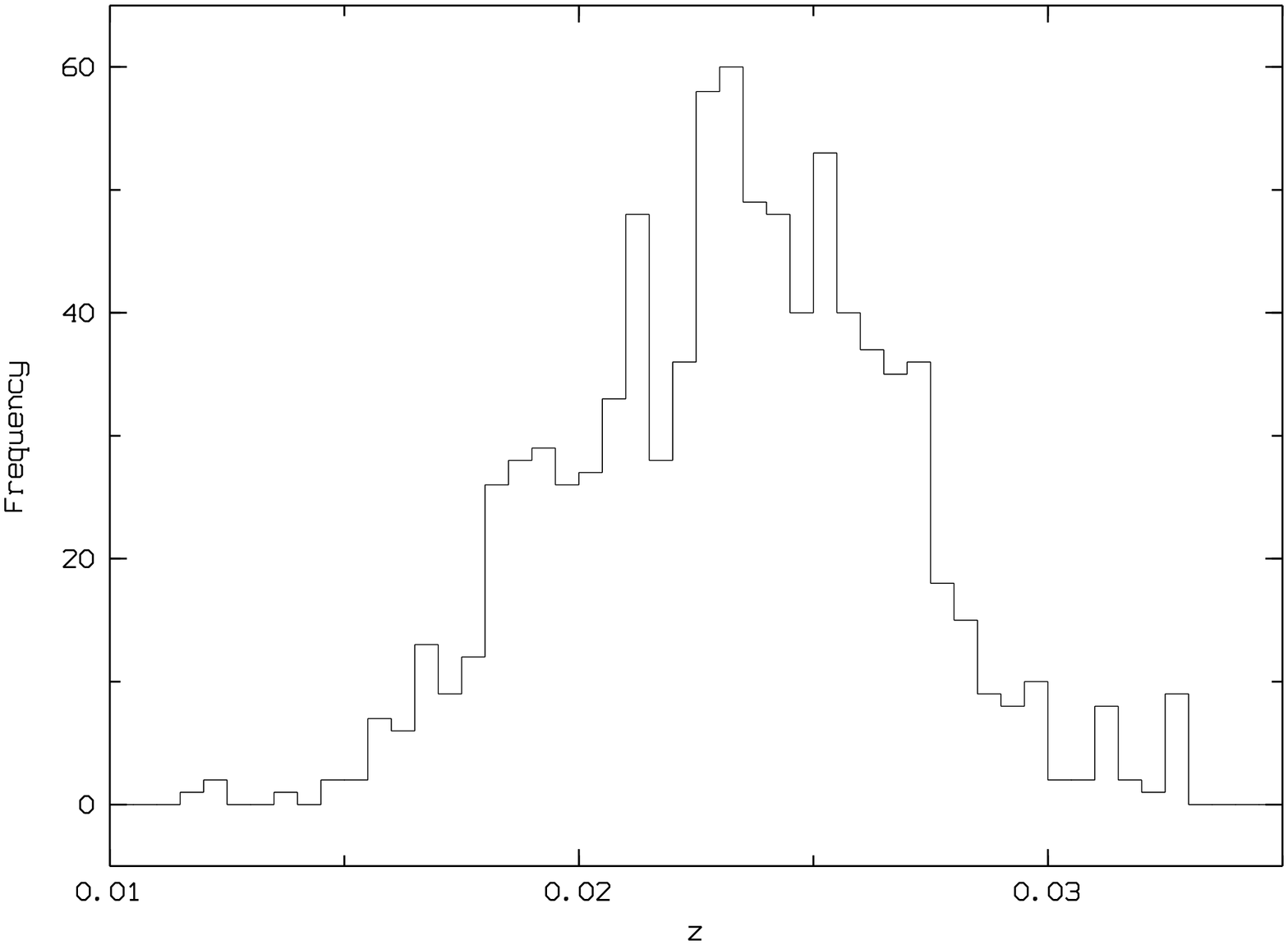,width=9cm,angle=270}}
\mbox{\psfig{figure=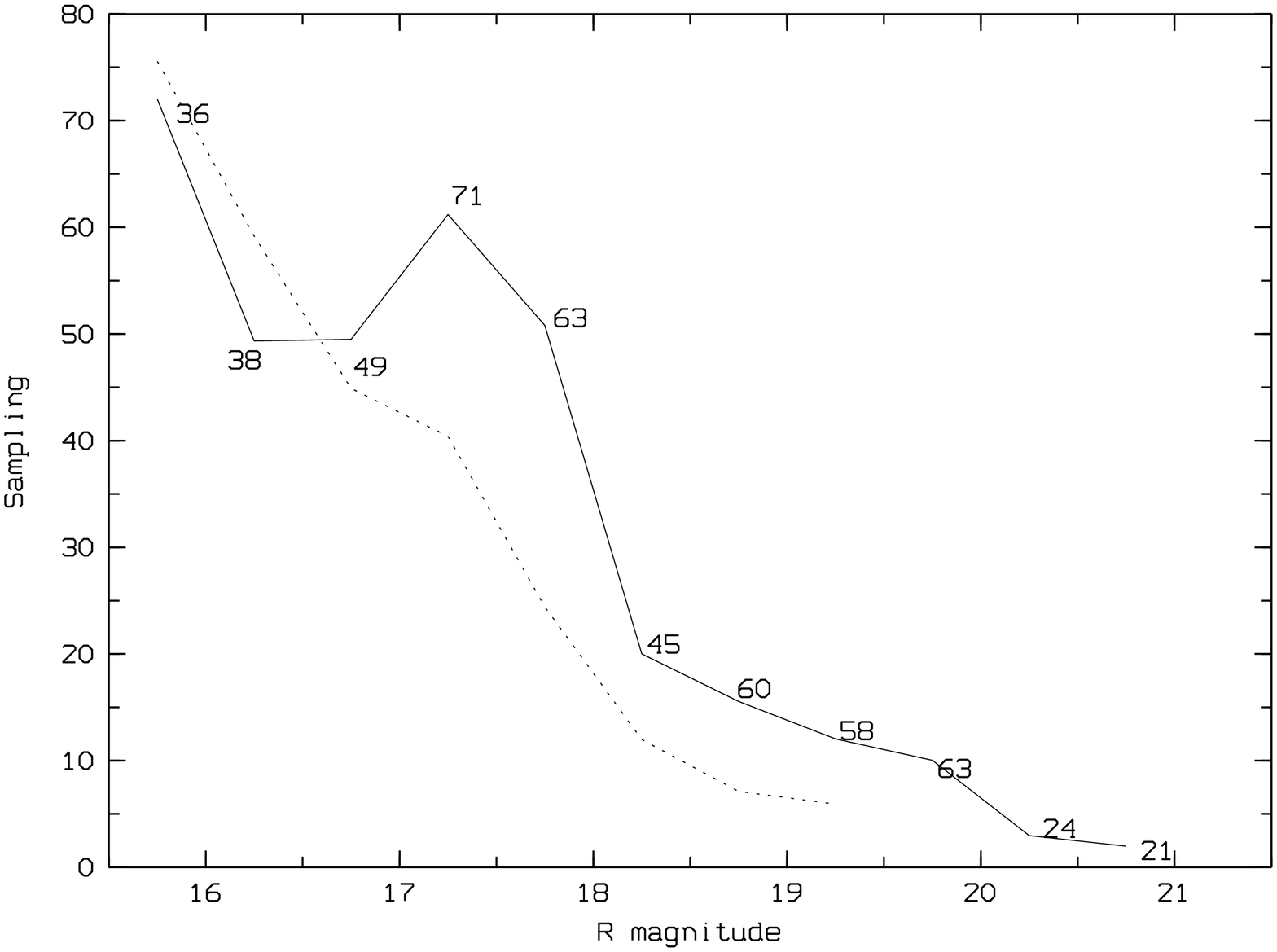,width=9cm,angle=270}}
\caption[]{Upper figure: redshift histogram of the sample between
z=0.01 and z=0.035.  Lower figure: completeness as a function of R
magnitude in the central area (solid line) and in the GMP area (dashed
line). The numbers indicated correspond to the total number of
galaxies with a measured redshift per magnitude bin.}
\label{fig:data}
\end{figure}

\begin{figure}
\mbox{\psfig{figure=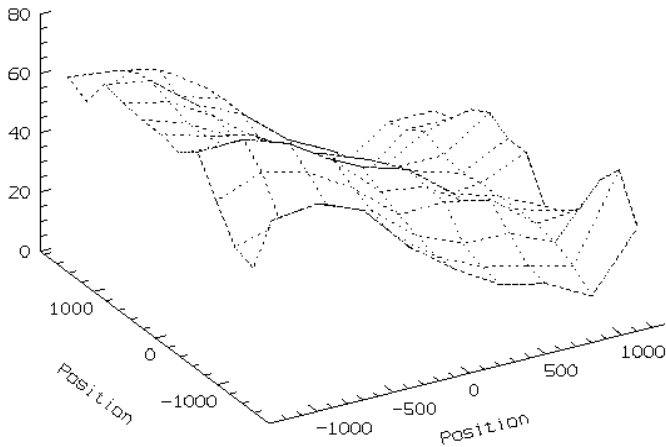,width=9cm,angle=270}}
\caption[]{Spatial redshift completeness over the CFHT imaging
area. Z-axis is the percentage of galaxies brigther than R=18 with a
measured redshift. The X and Y axes are the RA and declination 
given in arcsec relative to the GMP Coma cluster center. East is to
the left and West to the right.}
\label{fig:dataCFH}
\end{figure}

The velocities of galaxies in the region defined by GMP are taken from
the compilation of several literature sources used in Biviano et
al. (1996, hereafter B96), recently updated with the data of Rines et
al. (2001), Beijersbergen et al. (2002), Mobasher et al. (2003), and
Boselli (private communication). In total, we have 873 galaxies which
are considered cluster members (with velocities between 4000 and 10000
km/s in the GMP region). The sample is therefore significantly larger
than the one used by Colless \& Dunn (1996, hereafter CD96) and B96.
Galaxy magnitudes in each of these samples were scaled to the
R-band CFH12K system (see Adami et al. 2005b).

Fig.~\ref{fig:data} shows the redshift histogram of our catalogue
together with its completeness as a function of magnitude in the R
band. We defined two areas. The central area (solid line in
Fig.~\ref{fig:data}) is 42'$\times$55' centered on the Coma cluster
center and is close to the area surveyed with deep CFHT imaging data
(Adami et al. 2005b). The total area is the GMP area, with a limiting
magnitude of R=19.5 to avoid catalogue incompleteness.  Our
redshift catalogue is more than 50\% complete for R$<$18 and becomes
less than 20\% complete for R$>$18.5.

We also investigated the spatial completeness at R$<$18
(Fig.~\ref{fig:dataCFH}). We clearly see that the redshift completeness is
quite inhomogeneous over the CFHT field of view. The East areas are sampled at
about 60$\%$ while the West areas have a lower sampling close to 30$\%$. This
means that we should detect more easily the East substructures than the
West ones.

\begin{table}
\caption{Bright ($R<13$) galaxies in Coma. The columns are: (1)
galaxy name; (2) coordinates relative to the cluster center in
arcsec; (3) velocity in km~s$^{-1}$.}
\begin{tabular}{ccr}
\hline
Id & coordinates & $v$~~ \\
\hline
NGC 4789 & $4357, -3250$ & 8293 \\
NGC 4816 & $2826, -809$ & 6921 \\
NGC 4839 & $1873, -1702$ & 7389 \\
NGC 4841 & $1756, 1822$ & 6819 \\
NGC 4852 & $-4105, 4146$ & 5972  \\
NGC 4874 & $126, -39$ & 7189 \\
NGC 4889 & $-304, 23$ & 6472 \\
NGC 4911 & $-940, -647$ & 8026 \\
NGC 4921 & $-1340, -304$ & 5508 \\
NGC 4926 & $-1708, -1246$ & 7851 \\
NGC 4944 & $-3236, 774$ & 6942 \\
\hline
\end{tabular}
\label{tab:bright}
\end{table}

\begin{figure*}
\mbox{\psfig{figure=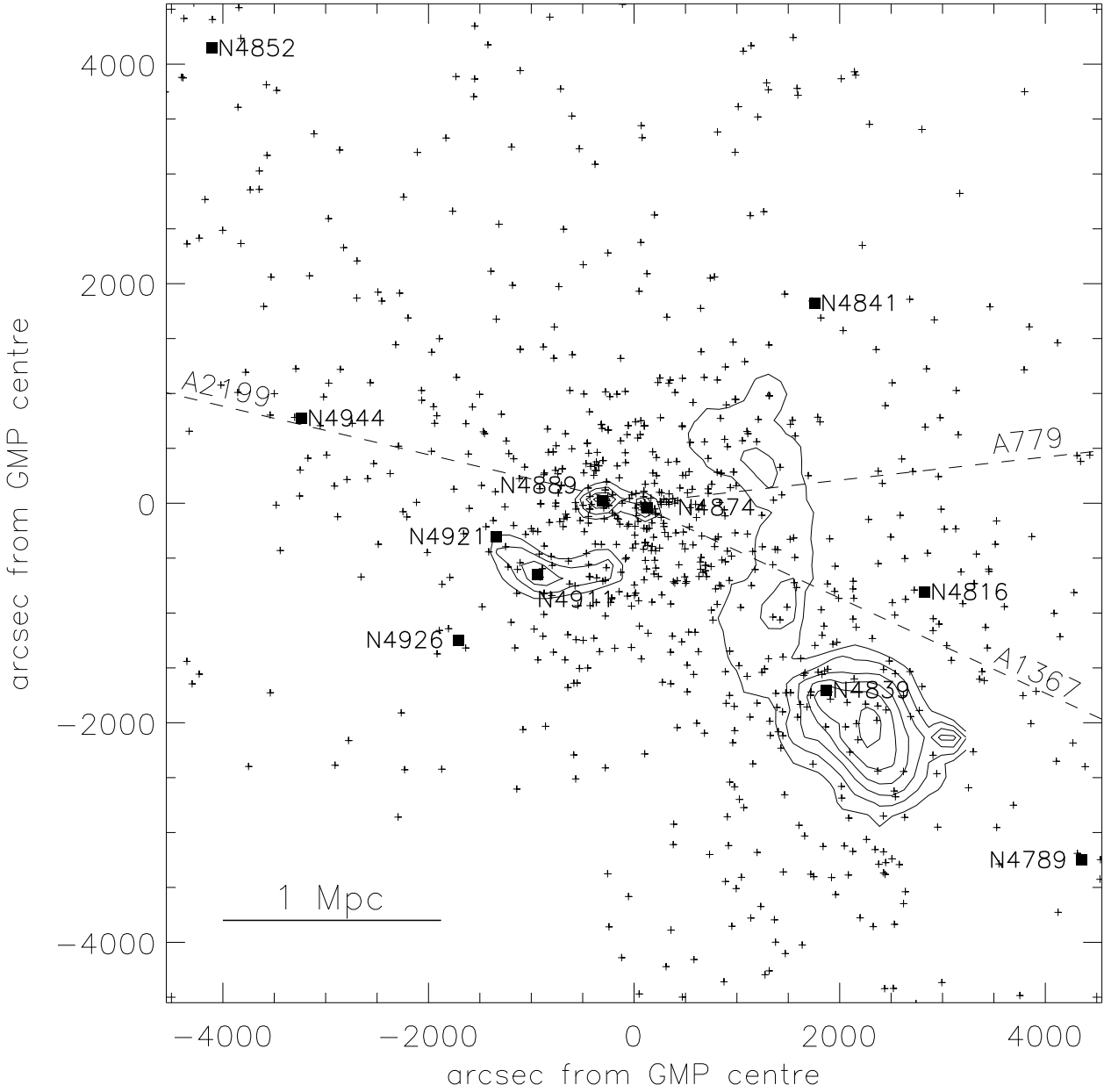}}
\caption[]{Spatial distribution of all cluster member galaxies with available
velocities. Filled squares identify galaxies with $R<13$, labelled
with their NGC number.  The
isocontours of the X-ray residuals over a $\beta$-model of N03 are
also shown. As in following figures, the X and Y axes are the RA and
declination, given in arcsec relative to the GMP Coma cluster center,
with North up and East to the left.  The solid segment represents
a length of 1 Mpc at the cluster distance.
The directions to three clusters
with redshift $z \leq 0.03$ in the vicinity of Coma are indicated.}
\label{fig:mapall}
\end{figure*}

The spatial distribution of all the cluster members (with measured
velocities) is shown in Fig.~\ref{fig:mapall}. The directions to three
clusters with redshift $z \leq 0.03$ and within a projected distance
smaller than 80~Mpc (at z=0.023) are also indicated (see also West
1998, Fig.~2).  We also reproduce in the same figure the isophotes
corresponding to the residuals identified by N03 after subtracting the
best-fit $\beta$-model from the X-ray emission of the whole Coma
cluster (see Fig.~2 in N03). In Table~\ref{tab:bright} we provide
names, coordinates, and velocities of galaxies with $R<13$,
shown as filled squares in Fig.~\ref{fig:mapall}.

\section{Substructure analysis}
\subsection{Application of the Serna \& Gerbal method}

In order to search for substructures in the Coma cluster, we first
applied the SG method to our redshift catalogue. This hierarchical
method allows to extract galaxy subgroups from a catalogue containing
positions, magnitudes and redshifts, based on the calculation of their
relative (negative) binding energies.  The method gives in output a
list of galaxies belonging to the selected group, as well as the
information on the binding energy of the group itself.

We use a M/L ratio in the R band of 200. This was derived from
the Coma cluster M/L ratio in the B band given by \L okas \& Mamon
(2003), using a colour $B-R=1.5$, typical of elliptical
galaxies.
We also checked that using a M/L ratio of 400 does not
change significantly the results. All groups are still detected and no
additional group is generated.  This range of M/L values ([200,400])
is also in good agreement with the values given by Carlberg et
al. (1996).

After each individual group was identified by the Serna \& Gerbal
method, we computed its mean velocity through the biweight estimator,
using the Rostat package (Beers et al. 1990).  Given the small number
statistics, and the contamination by cluster galaxies, the higher
velocity moments (such as the velocity dispersion) of these groups
have very large errors and are not reported here.

\begin{table}
\caption{Groups detected with the Serna \& Gerbal method. The columns
are: (1) group number; (2) mean group coordinates relative to the
cluster center in arcsec; (3) number of galaxies in the group; (4)
total $R$-band luminosity of group members in units of $10^{10}
L_{\odot}$; (5) mean group velocity and its 1$\sigma$ uncertainty, in
km~s$^{-1}$; (6) NGC identification numbers of the bright ($R<13$)
galaxies associated with the group.}
\begin{tabular}{rcrrll}
\hline
Id. & coordinates & N & $L_R$ & $\overline{v}$ & Bright gals. \\
\hline
 G1 & $-133, 53 $    & 19 & 22.9 & $6955 \pm 173$  & 4874, 4889 \\
 G2 & $1911, -1585$  & 10 & 24.2 & $7418 \pm 135$  & 4816, 4839 \\
 G3 & $1486, 1320$   &  5 &  9.8 & $7012 \pm 219$  & 4841 \\
 G4 & $-1334, -733$  & 11 & 14.4 & $7627 \pm 179$   & 4911, 4926 \\
 G5 & $-3662, 4$     &  6 &  9.0 & $6835 \pm 153$  & 4944 \\ 
 G6 & $4266, -2924$  &  4 &  8.9 & $8058 \pm 260$  & 4789 \\
 G7 & $-1617, 176$   &  8 & 19.7 & $5614 \pm 159$  & 4921 \\
 G8 & $470, -702$    &  3 &  2.9 & $6773 \pm 295$  & \\
 G9 & $715, -604$    &  3 &  2.6 & $5710 \pm 189$ & \\
G10 & $100, 704$     &  3 &  0.2 & $5636 \pm 52$  & \\
G11 & $2541, -3584$  &  4 &  6.4 & $6325 \pm 260$   & \\
G12 & $1687, -1079$  &  4 &  3.5 & $5897 \pm 155$  & \\
G13 & $1924, -1554$  &  4 &  2.2 & $4949 \pm 45$  & \\
G14 & $1544, -379$   &  3 &  2.0 & $6706 \pm 502$  & \\
G15 & $-1140, 3779$  &  7 & 10.0 & $7252 \pm 133$  & \\
G16 & $1787, 3772$   &  3 &  4.0 & $7498 \pm 70$ & \\
G17 & $4361, 418$    &  3 &  1.5 & $6710 \pm 269$  & \\
\hline
\end{tabular}
\label{tab:sample}
\end{table}

\begin{figure*}
\mbox{\psfig{figure=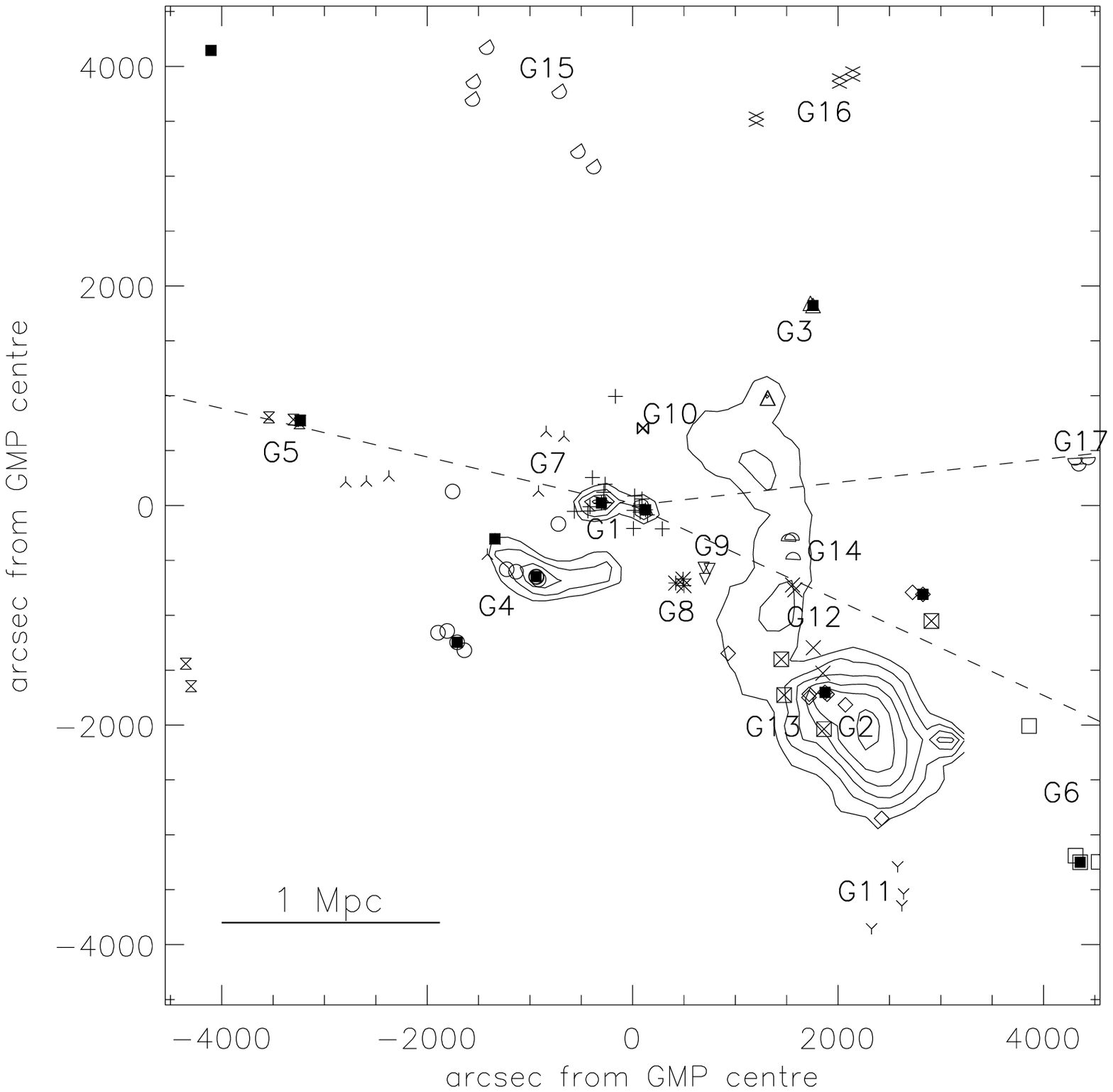}}
\caption[]{Spatial distribution of the galaxies belonging to the 17
groups detected by the SG method. Different symbols denote galaxies
belonging to different groups, except for bright galaxies ($R<13$)
which are plotted as filled squares. Groups are labelled with their
number. The solid segment represents a length of 1 Mpc at the cluster
distance.  The isocontours of the X-ray residuals over a $\beta$-model
of N03 are also shown. }
\label{fig:group}
\end{figure*}

\begin{figure}
\mbox{\psfig{figure=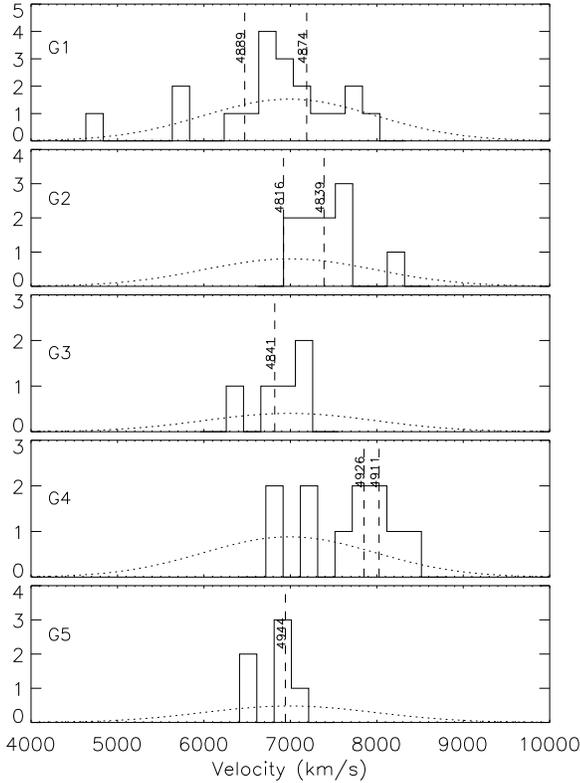,width=8cm}}
\caption[]{Velocity histograms of groups G1--G5 detected by the SG
method. The dashed line represents a gaussian with mean and dispersion
equal to the whole cluster mean velocity and velocity dispersion,
normalized to the total number of galaxies in each group. Vertical
dashed lines indicate the velocites of galaxies with $R<13$, labeled
by their NGC number, when present in the groups.}
\label{fig:hvel1}
\end{figure}

\begin{figure}
\mbox{\psfig{figure=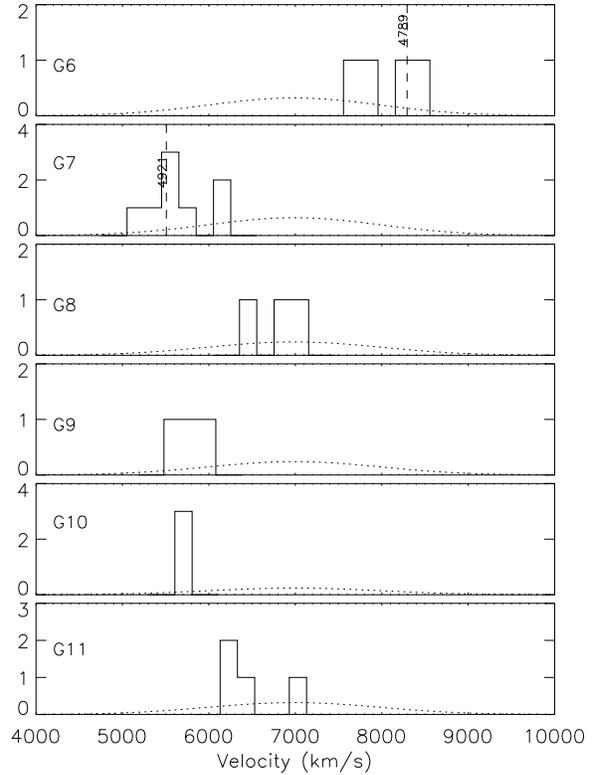,width=8cm}}
\caption[]{Same as Fig.~\ref{fig:hvel1}, but for groups G6--G11.}
\label{fig:hvel2}
\end{figure}

\begin{figure}
\mbox{\psfig{figure=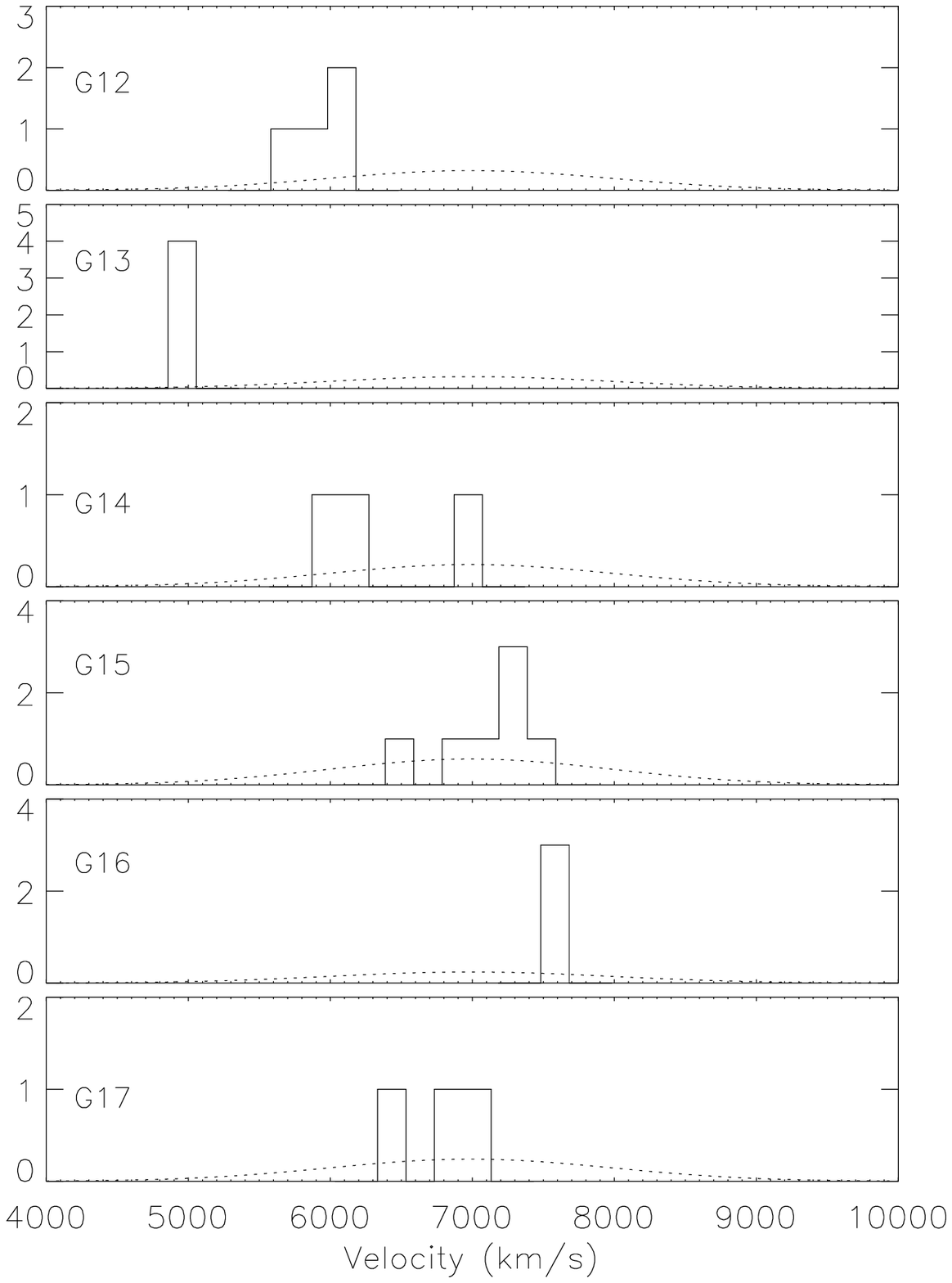,width=8cm}}
\caption[]{Same as Fig.~\ref{fig:hvel1}, but for groups G12--G17.}
\label{fig:hvel3}
\end{figure}

The SG method reveals the existence of 17 groups, whose
characteristics are given in Table~\ref{tab:sample}.  A map of Coma
with the positions of galaxies associated to the SG groups is
displayed in Fig.~\ref{fig:group}. The velocity histograms of the 17
groups are shown in Figs.~\ref{fig:hvel1}, \ref{fig:hvel2}, and
\ref{fig:hvel3}.  In those figures we also plot, as a reference, a
gaussian with mean and dispersion equal to the whole cluster mean
velocity and velocity dispersion, normalized to the total number of
galaxies in each group. In addition, the velocities of galaxies with
$R<13$, if present in the groups, are also indicated.  Here we briefly
describe the detected groups.

Group G1 is detected in the central region of Coma, and includes the
two central giant galaxies NGC~4889 and NGC~4874. It can be considered
the main cluster structure.

Group G2 is detected in the SW of Coma, and is centered around the
bright galaxy NGC~4839, both in coordinates and velocity space.  The
absolute value of its binding energy is even higher than the binding
energy of G1, which in fact, as we will see in Section~3.2, may itself
contain additional substructures. G2 is a very well known group, and
its properties have already been discussed by CD96.  The average
velocity of the group (and of its brightest member) suggests either an
infall motion from the foreground, or that the group has already
crossed Coma and is now escaping. The analysis of X-ray isophotes
clearly supports the former hypothesis (see Neumann et al. 2001). The
inclusion of NGC~4816 in this group is somewhat at odds with the
findings of M88, who found that NGC~4839 and NGC~4816 are located in
separate overdensity regions.

Group G3 is detected around the bright galaxy NGC~4841 in the NW Coma
region. Its mean velocity is close to the velocity of NGC~4841,
supporting a physical association between the group and its brightest
member. The binding energy of the group is smaller than that of G2 (in
absolute value) and of the order of that of G1. This binding energy is
also larger than that of all the other groups.  The mean velocity of
G3 is not far from that of the whole cluster, indicating slow motion
or a motion in the plane of the sky. This group was already apparent
in projection in the map of M88.

All the other groups have significantly lower binding energy (in absolute
values) than the first three.

Group G4 includes the very bright galaxies NGC~4911 and
NGC~4926 (with magnitudes $R<13$). Both galaxies are centered around
the mean group velocity.  This is in good agreement with the M88
analysis of the projected galaxy number density distribution,
suggesting the presence of a distinct group of galaxies around
NGC~4911.  The high velocity of the group relative to the cluster mean
implies an infalling motion from the foreground or, alternatively, an
outgoing motion, in which case the group should have already passed
through the cluster central regions. We discuss this point further in
\S~4.1. Note that this group was partially detected in X-rays by N03.

Three of the remaining groups are associated with very bright
galaxies, and hence have relatively high binding energies (in absolute
values): G5 contains NGC~4944, G6 contains NGC~4789, and G7 contains
NGC~4921.  In all three cases, the group mean velocity is close to the
velocity of the bright galaxy it contains. However, while the bright
galaxies of G5 and G6 are quite close to the group centre, this is not
the case for NGC~4921.  This suggests that NGC~4921 does not sit at
the bottom of the group potential well.  At variance with G6 and G7,
G5 has a mean velocity similar to that of the whole cluster, and hence
is either moving slowly, or has its velocity vector aligned with the
plane of the sky. This group was already visible as an enhancement of
the galaxy density in the map of Mellier et al. (1988).  It is
interesting to note that both G5 and G7 lie along the direction
connecting the cluster centre to the neighbouring cluster Abell~2199, and
G6 lies along the direction connecting the cluster centre to
Abell~1367.

Of the remaining groups, three (G8, G9 and G10) are located very close
to the cluster centre, while groups G11, G12, G13 and G14 are close to
the NGC~4839 group but with quite different velocities, and hence
probably physically unrelated to the NGC~4839 group.  G12 and G14 are
close to one of the X-ray peaks detected by N03, the southern part of
the so-called extended western substructure.

Groups G15 and G16 are located north of the Coma
cluster. Interestingly, {both groups are found approximately along the
direction connecting the centre of Coma to G2, and share a similar
velocity to that of G2.  Hence we could speculate that they have}
infallen into Coma along the same direction, perhaps from the same
large-scale structure filament that connects Coma to Abell~1367 (see
Fig.~2 in West 1998). Similarly, it is interesting to remark that
G17, a loosely bound group of three galaxies, is located in the
direction of another neighbouring cluster, Abell~779.

\subsection{Groups corresponding to X-ray surface brightness enhancements}

\begin{figure*}
\mbox{\psfig{figure=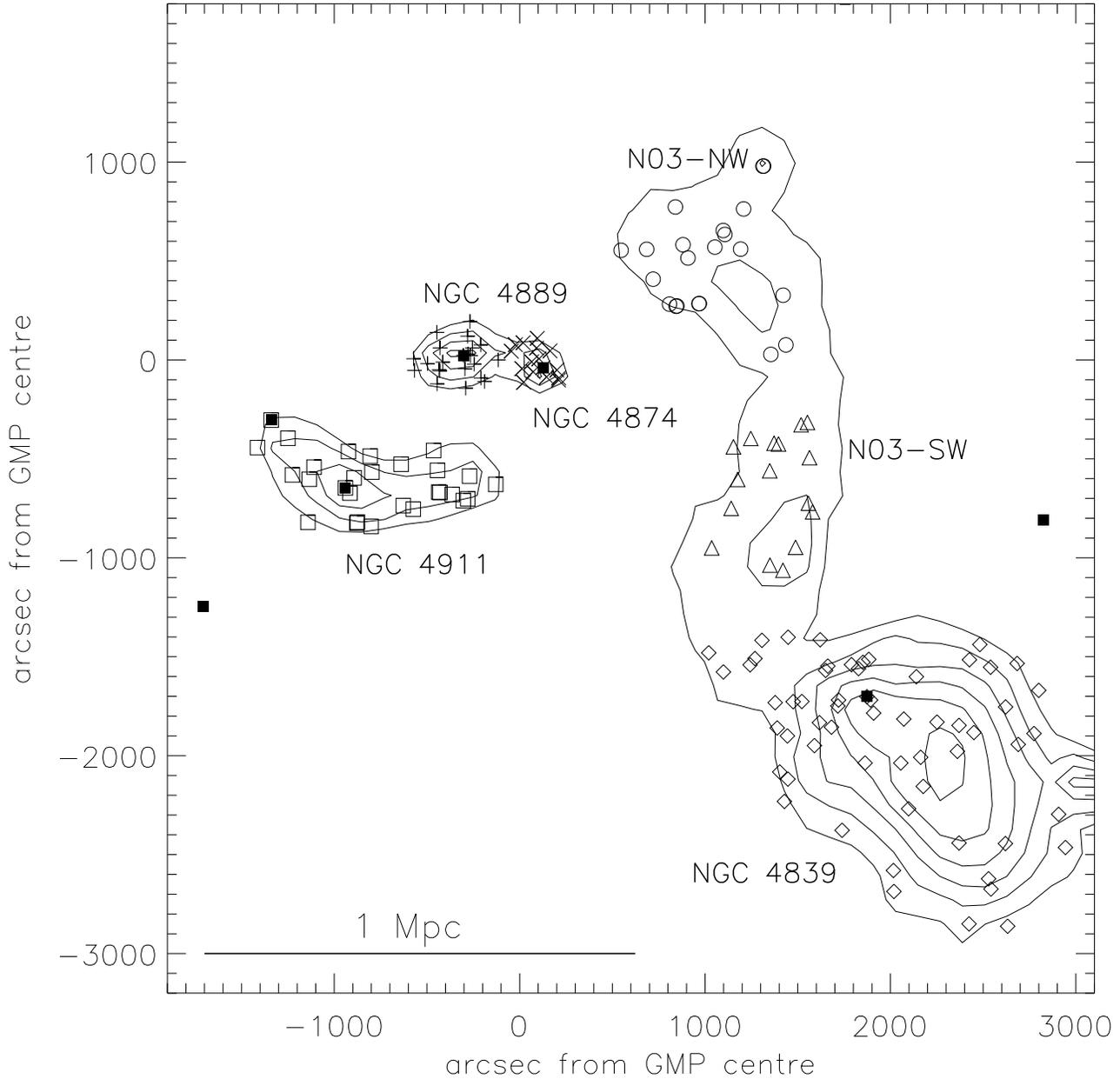}}
\caption[]{Spatial distribution of the galaxies associated with
6 peaks of the X-ray surface-brightness residuals. Different symbols denote
galaxies belonging to different groups, except for bright galaxies
($R<13$) which are plotted as filled squares. The solid segment represents
a length of 1 Mpc at the cluster distance.
 Also shown are the isocontours of the X-ray
$\beta$-model residuals of N03. }
\label{fig:residuals}
\end{figure*}

Additional information on the subclustering properties of the Coma
cluster comes from the X-ray band observations of N03. These authors
have identified residuals in the X-ray surface brightness of Coma,
after subtraction of the best-fit $\beta$-model. Significant residuals
were found around the bright galaxies NGC~4889, NGC~4874, NGC~4911,
NGC~4921, NGC~4839, plus an additional, elongated residual to the west
(see Fig.~2 in N03 and Fig.~\ref{fig:residuals}).  With the SG
method we have detected a single group associated to the two distinct
X-ray groups around NGC~4889 and NGC~4874, three groups related to
NGC~4911, NGC~4921, and NGC~4839, and another two groups associated
with the southern part of the extended western X-ray residual (groups
G12 and G14). An exact correspondence between X-ray and optically
detected groups is not, however, to be expected. Significant
displacements between the gas and galaxy components of an infalling
group are likely to occur as a consequence of the Intra-cluster
medium acting to slow down the group diffuse gas, while galaxies are
freely streaming. This displacement is actually visible in the group
around NGC~4839, with the brightest galaxy of this group clearly ahead
of the group gas which is trailing behind in the group infall motion
towards the cluster center (see Neumann et al. 2001).

\begin{figure}
\mbox{\psfig{figure=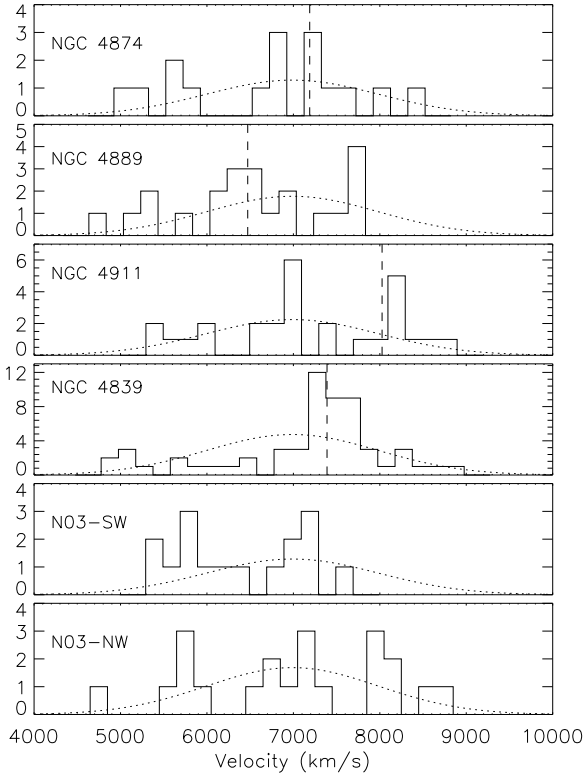,width=8cm}}
\caption[]{Same as Fig.~\ref{fig:hvel1} for the six subsamples of
galaxies located in the isocontours of the X-ray surface-brightness
residuals.}
\label{fig:hvel4}
\end{figure}

In any case, we deemed it interesting to look for signatures of the
X-ray subgroups in the galaxy component, besides the Serna \& Gerbal
detections. For this, we selected galaxies within the isocontours of
the X-ray surface brightness residuals, as shown in
Fig.~\ref{fig:residuals}, and we analyzed their velocity distributions
(these are plotted as histograms in Fig.~\ref{fig:hvel4}).  Given that
the elongated X-ray western residual has two main peaks, we have
considered them separately.  The results of our analysis are given in
Table~\ref{tab:residuals}, where we give in particular the probability
that the velocity distribution of the selected galaxies in the group
is drawn from the velocity distribution of the cluster as a whole,
according to a Kolmogorov-Smirnov test. The fact that the velocity
distribution of a subsample is not a random draw from the whole
cluster velocity distribution, makes this subsample a likely physical
group.

\begin{table}
\caption{Kinematical properties of galaxy subsamples selected in six
regions corresponding to the X-ray residuals emission over a
$\beta$-model of N03.  The
contents of the columns are: (1) subsample id. name; (2) biweight mean
coordinates of the galaxies in the subsample, relative to the cluster
center in arcsec; (3) number of galaxies in the subsample; (4) mean
velocity of the galaxies in the subsample, and its 1$\sigma$
uncertainty, in km~s$^{-1}$; (5) Kolmogorov-Smirnov probability that
the velocity distribution of the galaxies in the subsample is a random
draw from the velocity distribution of the whole cluster.}
\begin{tabular}{ccrlr}
\hline
Id. & coordinates & N & $\overline{v}$ & KS prob. \\
\hline
NGC~4874 & $92, -11$ & 16 & $6753 \pm 271$ & 0.74 \\
NGC~4889 & $-334, -6$ & 22 & $6568 \pm 204$ & 0.12 \\
NGC~4911 & $-764, -622$ & 28 & $7165 \pm 130$ & 0.32 \\
NGC~4839 & $1977, -1859$ & 59 & $7459 \pm  97$ & $<0.01$ \\
N03-SW   & $1374, -626$ & 16 & $6331 \pm 201$ & 0.02 \\
N03-NW   & $1021, 490$ & 21 & $7013 \pm 258$ & 0.69 \\
\hline
\end{tabular}
\label{tab:residuals}
\end{table}

Only two subsamples have velocity distributions that are significantly
different from that of the whole Coma cluster: the one corresponding
to the substructure around NGC~4839, and another one corresponding to
the southern part of the extended western structure of N03
(N03-SW). The velocity range of the last one is in good agreement with
G12 and G14. We therefore re-discover the NGC~4839 group as
well as the groups G12 and G14. Note that N03 have not been able to
find any galaxy overdensity associated with the western elongated
X-ray residual, while this is now possible thanks to the added
kinematical information.

Note that the group we detected including NGC~4911 is not
recovered by this method when limiting ourselves to the X-ray emission area
only (around NGC~4911), perhaps because of its central position
which makes it subject to substantial contamination by members of the
Coma cluster itself.

\section{Discussion}

\subsection{Detected groups}

Using the SG method applied to the projected phase-space
distribution of galaxies in the Coma cluster, we have identified 17
dynamically bound groups. Four of them are also identified in X-rays
by N03, but many others have not been previously identified, and
enable us to deepen our understanding of the complex structure of the
Coma cluster.

N03 have detected two independent groups in the central region of
Coma, one associated with NGC~4889, and another with NGC~4874. Our
analysis suggests instead that these two galaxies are part of the same
group, which could actually be identified as the core of the Coma
cluster itself.  A hint that the two galaxies actually do lie in
distinct substructures comes from the visual inspection of the
velocity histograms of the X-ray selected galaxy subsamples (see
Fig.~\ref{fig:hvel4}), where NGC~4874 and NGC~4889 seem to lie at the
peaks of the velocity distributions of their surrounding galaxies.
The significance of these peaks is however marginal. Many other bright
galaxies do however happen to be located at (or close to) the
barycenters of the groups which they reside in. Hence our findings
confirm the original claim of M88.

The NGC~4911 group was shown by N03 to be a low mass
group of galaxies, with a mass of the order of $5 \times 10^{12}
M_\odot$.  N03 suggested that this group is currently experiencing ram
pressure stripping. The fact that the group still has a sufficently
high Intra-group gas density to be detectable in X-rays, argues
against the group having already crossed the cluster core. Indeed,
small groups are easily disrupted, and their X-ray gas is broken up
with respect to the galaxy component, after they cross the cluster
core (e.g. Tormen et al. 2004).  Its average velocity then implies the
group is currently infalling into the cluster. The shapes of the group
X-ray isophote, and the lack of a shock-heated gas region near the
group location, argue however against a radial infall. More likely,
the group--cluster collision has a non-zero impact parameter, and the
group is spiralling into the cluster. 

Note however, that our analysis does not confirm the association of
NGC~4921 with the NGC~4911 group. This is because the difference in
velocities between the two galaxies is $\sim 2500$ km~s$^{-1}$ (see
Table~\ref{tab:bright}; note that the value reported by N03 for the
velocity of NGC~4921 is incorrect -- G.~Gavazzi, private
communication), and because NGC~4921 is found to be associated with
another independent group (G7).

The group around NGC~4839 (G2) is the most strongly bound of all
detected groups, even more than the central structure (G1) which is in
fact identified with the cluster itself. G1 is less bound than G2
perhaps because the original compactness of G1 has been reduced by a
group-cluster collision (e.g.  that with the presumed NGC~4889 group,
see the following section). On the other hand, the very high binding
energy (in absolute values) of the NGC~4839 group strenghtens the
conclusions already reached by Neumann et al. (2001) that this group
has not yet crossed the cluster core.

Finally, it is worth mentioning the groups G12 and G14, that are
located close to the southern part of the western substructure
identified by N03 in X-rays. While N03 did not actually distinguish
between the northern and southern parts of the western X-ray
extension, the extension is clearly two-peaked. The association of the
southern peak with a galaxy group, and the lack of any galaxy group
associated with the northern peak, suggest the two X-ray peaks are
physically distinct features. We come back to this issue in the
following section.

\subsection{Central region: which is the colliding galaxy?}

The two central galaxies have individual X-ray haloes, as found by
N03, and maybe they have small groups dynamically associated with
them (although we still fail to identify them unambiguously):  

 - The
galaxy distribution around NGC~4874 is clearly not different from the
whole Coma cluster galaxy velocity distribution (see
Table~\ref{tab:residuals}). 

 - By comparison, the Kolmogorov-Smirnov
probability of the galaxy velocity distribution around NGC~4889 to be
similar to the whole Coma cluster distribution is not very high,
marginally indicating a galaxy group with a more or less independent
status surrounding this bright galaxy.  

It is also remarkable that the
velocities of the two main Coma cluster galaxies are offset with
respect to the cluster mean, by $\sim -500$ (NGC~4889) and $\sim +200$
(NGC~4874) km~s$^{-1}$. Such velocity offsets are not common among the
dominant galaxies of clusters (Gebhardt \& Beers 1991) and probably
indicate deviation from dynamical relaxation. Maybe one of the two
galaxies was originally sitting at the bottom of the Coma cluster
potential well, but was later displaced and gained kinetical energy
through the collision with the group associated with the other galaxy,
as originally suggested by CD96. The question now is which is
the colliding galaxy?

\subsubsection{NGC~4874 is the colliding galaxy?}

On the one hand, NGC~4874 could be the colliding galaxy with an
appropriate northward motion as suggested for example by Gurzadyan
$\&$ Mazure (2001) or Adami et al. (2005a). This is supported by the
presence of a possible X-ray shock front north of this galaxy and by
the later galaxy content and high binding energy of the group
surrounding this galaxy detected by Gurzadyan $\&$ Mazure (2001)
(compared to the NGC~4889 group).  In this scenario, we would expect,
however, to also detect such a residual galaxy group around NGC~4874,
and contrary to Gurzadyan $\&$ Mazure (2001), this is clearly not the
case with the Serna-Gerbal method or using the Kolmogorov-Smirnov test
(see Table~\ref{tab:residuals}). 

This group has perhaps already been
diluted, but we still need to explain the group possibly present 
around NGC~4889 (see beginning of discussion). The colliding NGC~4874 
scenario indeed implies that
NGC~4889 has been in the Coma cluster longer and this makes 
difficult to explain why it would still show signs of a surrounding
group contrary to NGC~4874. We could indeed expect the group
surrounding NGC~4889 to be more diluted than the NGC~4874 one if it
has been present in Coma for a longer time. 

Moreover, Adami et
al. (2005a) argued that diffuse light sources around NGC~4889 could
already have been re-accreted by this galaxy, while still present
around NGC~4874, showing that NGC~4889 is the original cD of the Coma
cluster. However, recent numerical simulations by Napolitano et
al. (2003) show that diffuse light sources remain unrelaxed even at
z=0 with a highly non-Gaussian velocity distribution. Hence, should
NGC~4889 have been originally sitting at the bottom of the cluster potential
well, some diffuse light around it should still be visible.

\subsubsection{NGC~4889 is the colliding galaxy?}

On the other hand, N03 has suggested that NGC~4889 is the colliding
galaxy, associated with the gas that has been stripped by ram pressure
and is now visible as the western extension in N03's residuals
map. N03's interpretation originates from the fact that no other
galaxy overdensity seems to be associated with the western X-ray
structure. 

However, we do identify a galaxy group in the southern
region of the western emission, as a kinematically distinct entity
from the whole cluster (N03-SW; or perhaps two: G12 and
G14). Remarkably, the mean velocity of this group is similar to that
of NGC~4889, but it is hard to see how the gas, if stripped from
NGC~4889, would now appear to be related with the galaxies in
N03-SW. However, the northern part of the X-ray western structure of
N03 does not correspond to any galaxy group, and could still be
stripped gas from the infallen NGC~4889 group. If NGC~4889 is the
colliding galaxy, we can then draw the following scenario.  N03-SW is
a real group, currently infalling (from the back, given its radial
velocity) into Coma, and producing the X-ray temperature enhancement
visible in Fig.~3 of N03. The northern part of the western extended
X-ray structure is not related to N03-SW, and is in fact residual gas
from a group associated with NGC~4889. In this case, this group
would have been accreted onto Coma flowing along the direction
connecting Coma to the neighbouring cluster Abell 779.  The shift
between the galaxy and the group gas must have occurred some 0.5--1
Gyr ago (given typical Coma galaxy velocities and the distance between
the gas and the galaxy). 

The impact between the NGC~4889 group and the
main cluster could have displaced NGC~4874 from the bottom of the
cluster potential well and, at the same time, disrupted the cluster
cool core, if originally present -- Coma is a well known
counter-example of a cooling-core cluster, its X-ray emission being
unusually flat near the centre (e.g. Fabian 1994). The induced motion of
NGC~4874 would be approximately northward to be in agreement with
Adami et al. (2005a). 

This scenario does not explain, however, why the
NW X-ray residual would be related to NGC~4889 while we still see
another X-ray residual directly around this galaxy (N03). Perhaps, one
may think of a multiphase gas component in the NGC~4889 group, one
component associated to the group as a whole and another, colder
component, more strictly bound to the giant galaxy (as the one shown by
Vikhlinin et al. 2001 with a temperature of 1 to 2 keV).
\\
\\
In conclusion, the two above described scenarios are both probably too
simplistic and fail to fully account for the complex observational
phenomenology, although the NGC~4889 colliding galaxy hypothesis seems more
appealing.

\subsection{Group infalling process}

\begin{figure}
\mbox{\psfig{figure=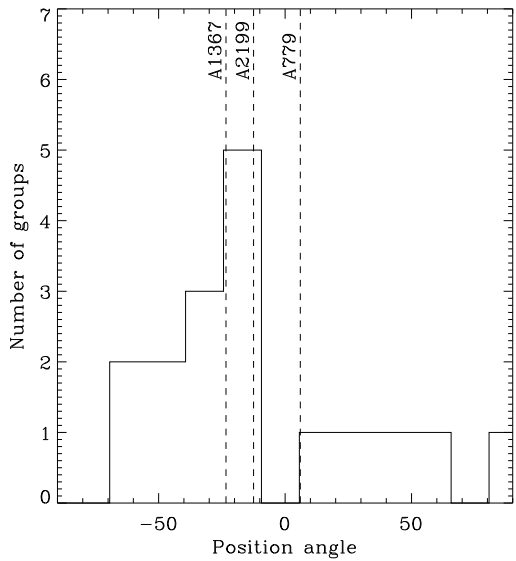,width=8cm}}
\caption[]{The distribution of position angles of the 17 detected
groups. Dashed lines indicate the position angles of the directions
to three neighbouring clusters.}
\label{fig:posang}
\end{figure}

A remarkable feature of the detected groups is that most of them lie
along the main directions of the large scale structure surrounding
Coma, as traced by the direction to three relatively nearby clusters,
Abell~779, Abell~1367, and Abell~2199 (see Fig.~\ref{fig:mapall}; see
also West 1998). This is clearly seen in Fig.~\ref{fig:posang}
where we show the distribution of the position angles of the 17
groups, as well as the directions to the three neighbouring clusters.
Our finding then strengthens the idea that Coma is currently accreting
groups from the surrounding large scale structure.

The accretion needs not to be radial, however. We already discussed
the case of the NGC~4911 group, which does not seem to be infalling
radially onto the center of Coma (see \S~4.1). B96 argued for a
non-radial infall of the NGC~4839 group as well, in order to explain
the coincidence of the phase-space location of galaxies in this group
and the E+A population discovered by Caldwell et al. (1993).
Tangential orbits would not be surprising, as groups infalling along
radial orbits are selectively destroyed (Gonz\'alez-Casado et
al. 1994; Taylor \& Babul 2004) and we preferentially identify groups
on tangential orbits, since they survive longer. Longer survival times
help to explain why most of our detected groups lack X-ray emission,
since groups orbiting the central regions of Coma, but not crossing
its core, can have their X-ray gas stripped by ram pressure, and still
survive disruption.

Biviano et al. (2002), analysing the ESO Nearby Abell Cluster
Survey (ENACS, Katgert et al. 1996, 1998) data-set, have in fact shown
that galaxies in substructures are characterized by a lower velocity
dispersion than other cluster galaxies. This finding was later
interpreted by Biviano \& Katgert (2004) as evidence for tangential
orbits. On the other hand, the high-velocity dispersion of late-type
galaxies was found to be an indication of (mildly) radial
orbits. Hence, in order to better understand which kind of orbits
characterize the infalling groups, we look at their kinematics. For
the 17 Coma cluster groups we find a mean velocity of $6675 \pm 220$
km~s$^{-1}$, and a velocity dispersion of $872_{-140}^{+165}$
km~s$^{-1}$, both consistent with the corresponding values for the
whole Coma cluster ($6989 \pm 34$ km~s$^{-1}$ and $973_{-50}^{+47}$
km~s$^{-1}$). Since galaxies in Coma move on nearly isotropic orbits
(\L okas \& Mamon 2003), it is tempting to conclude that the same is
true of the Coma groups, but the uncertainties are currently too large
to exclude either moderately tangential or moderately radial orbits.

Assuming a maximum survival time of 2--3 Gyr for our groups
(Gonz\'alez-Casado et al. 1994), we can try to estimate the mass
accreted onto Coma since $z \sim 0.2$--0.3. This is likely to be a
lower limit estimate, since disrupted groups would not be counted, and
the survival time of some of the detected groups is likely to be
shorter.  We do not have a direct estimate of the mass of our
groups, except for the group around NGC~4839, which has an estimated
mass of $8.6 \times 10^{13} M_{\odot}$ (CD96), and for the group
around NGC~4911 (which we detect only in association with NGC~4926
using the SG method), which has an estimated mass of $0.5 \times
10^{13} M_\odot$ (N03). Both groups are associated with X-ray emitting
gas, at variance with most other groups. Hence, their masses are
likely to be larger than the average group masses. If we adopt
the geometrical mean of these two estimates, $2.1 \times 10^{13}
M_{\odot}$, as the typical mass of the detected groups in the Coma
cluster, we obtain a total mass for the 17 detected groups of $3.7
\times 10^{14} M_{\odot}$.  Alternatively, we could adopt the average
mass of groups seen in large-scale surveys, $2.8 \times 10^{13}
M_{\odot}$ (Ramella et al. 1989), yielding a total mass of $4.8 \times
10^{14} M_{\odot}$.  Finally, we can convert the group luminosities
into masses, using the mass-luminosity relation recently derived by
Popesso et al. (2005).  Through their relation we obtain a total mass
of $1.5 \times 10^{14} M_{\odot}$ for the 17 groups. We can thus
estimate that Coma has accreted between 1.5 and $4.8 \times 10^{14}
M_{\odot}$ since $z \sim 0.2$--0.3.
This corresponds to $\sim 10$--30\% of the total cluster mass out to
the virial radius (we adopt here the estimate of \L okas \& Mamon
2003, $1.4 \times 10^{15} M_\odot$), in fair agreement with the
expectations from hierarchical models of cluster galaxy formation
(e.g. Tormen et al.  1997). Put in another way, if the mass accretion
rate has not changed significantly since a much higher redshift, the
estimated accretion rate of (0.4--2.0)$\times 10^{14} M_\odot$~Gyr$^{-1}$
implies that Coma could have formed by accretion of groups from the
surrounding large scale structure in less than 14 Gyr, again in
agreement with the hierarchical $\Lambda$CDM models (e.g. Evrard et
al. 2002)

\section{Summary and conclusions}

With the aim of searching for a possible scenario to explain the
buildup of the Coma cluster, we have searched for new substructures by
applying the SG hierarchical method to a catalogue of about 900 galaxy
redshifts and magnitudes. We have also taken into account recent
results based on XMM-Newton images, which show residual emission over
a smooth $\beta$-model (N03).

In total, 16 groups were identified in addition to the central main
cluster core. Several of these groups had not been previously
identified. Most of them are distributed along the directions towards
the three neighbouring clusters Abell~779, Abell~1367 and Abell~2199,
suggesting that they may have been accreted from the surrounding large
scale structure. Accretion onto the cluster is however unlikely to
occur on purely radial orbits, or these (not very massive) groups
would be easily disrupted by the tidal field.

We estimate that Coma has accreted $\sim 1/5$ of its mass in the form of
groups since $z \sim 0.2$--0.3. Such an accretion rate is consistent
with the Coma cluster having been built hierarchically over a Hubble
time in the currently favoured $\Lambda$CDM model.

Even with 900 galaxy redshifts, our knowledge of the Coma cluster
dynamics is still limited to rather bright galaxies. In order to
explore previous suggestions of a different dynamical status of bright
and faint galaxies (B96), a spectroscopic follow-up of Coma galaxies
at fainter magnitudes, requiring an 8~m class telescope, is needed.

\begin{acknowledgements}
We thank the referee, A.~Serna, for useful and constructive comments.
We also thank A. Boselli, M. Colless, and R.C. Nichol for
providing us with additional redshift measurements, and G. Gavazzi for
useful discussion.  We are grateful to D. Neumann for providing us
with an electronically readable version of the X-ray residuals map.
C.A. and F.D. thank the PNG (CNRS) for financial support. A.B. acknowledges
the hospitality and financial support of the Laboratoire
d'Astrophysique de Marseille, of the Institut d'Astrophysique de
Paris and of GDRE EARA (CNRS).  \\
\end{acknowledgements}

\end{document}